\documentclass[11pt,preprint]{aastex}
\usepackage{natbib}
\begin{document}
\title{The compositions of Kuiper belt objects}

\author{Michael E. Brown}
\affil{Division of Geological and Planetary Science, California Institute of Technology, Pasadena, CA 91125}

\slugcomment{Annual Review of Earth and Planetary Sciences, In Press}

\begin{abstract}
Objects in the Kuiper belt are small and far away thus difficult
to study in detail even with the best telescopes available at earth.
For much of the early history of the Kuiper belt, studies of the
compositions of these objects were relegated to collections of
moderate quality
spectral and photometric data that remained difficult 
to interpret.  Much early effort was put into simple correlations
of surface colors and identifications of spectral features,
but it was difficult to connect the observations
to a larger understanding of the region. The last decade, however, 
has seen a blossoming in our understanding of the compositions 
of objects in the Kuiper belt. This blossoming is a product of the
discoveries of larger -- and thus easier to study -- objects, continued 
dedication to the collection of a now quite
large collection of high quality
photometric and spectroscopic observations, and continued work
at the laboratory and theoretical level. Today we now know 
of many processes which affect the surface compositions of
objects in the Kuiper belt, including
atmospheric loss,
differentiation and cryovolcanism, radiation processing, the
effects of giant impacts, and the early
dynamical excitation of the Kuiper belt. We review the large quantity of
data now available and attempt to build a comprehensive framework
for understanding the surface compositions and their causes. In contrast
to surface compositions, the bulk compositions of objects in the 
Kuiper belt remain poorly measured and even more poorly understood,
but prospects for a deeper understanding of the formation of the 
the outer solar are even greater from this subject.
\end{abstract}

\section{Introduction}
The region of the solar system 
beyond the planets is populated by a vast 
swarm of objects in what is called the Kuiper belt. The objects
in this swarm -- Kuiper belt objects (KBOs) -- formed on
the outer edge of the planetary system where nebular densities were
too low and accretional timescale too high to allow objects to
accumulate into a single dominant planet, as happened throughout
the planetary system. Instead, in the region beyond the planets, 
accretion proceeded up to objects at least as large as the 2700 km diameter
of Triton but then stalled as dynamical excitation of the region
caused collision velocities to be too high to allow continued accretion. 
Today, encounters between objects in the outer solar system lead
to erosion to smaller diameters.

Much of we have learned from the Kuiper belt has come from 
study of the dynamics of these bodies \citep[i.e.][]{2008ssbn.book..275M}. Understanding
the dynamical state of a KBO is straightforward: after an
object is discovered, a few observations spaced over several years
are usually sufficient to define a precise orbit for the object
\citep{2000AJ....120.3323B}. Since their formation
KBOs have been moved around by the gravitational tugs and perturbations
of the giant planets, and their current locations preserve a record
of these billions of years of interactions. Our current understanding
of the rearrangement and migration of the giant planets is based 
largely on this record. 

In principle, the objects in the Kuiper belt also contain a wealth of
information on the chemical conditions in the earliest solar system.
The materials in these objects have, for example, avoided much of
the high temperature and pressure processing to which materials in
the inner solar system have been subjected. 
In practice, however, understanding the detailed composition
of KBOs in their native environment has proved difficult.
Most objects are small, far away, and thus difficult to study
even with the largest telescopes available. For the small number
of objects large enough for detailed compositional study to
be feasible, only the surface chemistry, which is clearly
not representative of the entire body, is accessible to observation.
For detailed understanding of ices and minerals incorporated into
bodies in the early outer solar system, short-period comets, 
which are generally derived from the Kuiper belt, are much better
targets for study. 

For the first decade of the study of compositions of KBOs, most effort
was put into simple cataloging and classifying the surfaces, with
little insight into underlying causes. The past decade, however, has
seen a maturation of both the data sets available and the 
physical and chemical insights into their meaning. 
The chemical composition of KBOs serves a tracer of the
physical and chemical 
processes to which these bodies were subjected  at formation
and in the time period since. Using studies of surface composition
we have gained insight into atmospheric processes,
differentiation and cryovolcanism, radiation processing, the
effects of giant impacts, and the early
dynamical excitation of the Kuiper belt. 

In this review we attempt to formulate a coherent picture
explaining the compositions of Kuiper belt objects. 
We show that KBO surface composition divides most strongly
by size, with large, medium, and small objects having
distinct surface types, and we propose physically and
chemically plausible hypotheses for all of these surfaces.
We attempt to formulate similar hypotheses for
the bulk compositions of KBOs, but we find no satisfactory
hypothesis that fits the current data. We end with 
some thoughts on important open questions that would best
allow progress to be made in these areas.

\section{The largest Kuiper belt objects}
\subsection{Volatiles}
Spectroscopic study of some of the earliest discovered KBOs 
revealed surfaces that were either spectrally featureless or
contained only small amounts of absorption due to water ice. 
These early spectra of relatively small objects were in marked
contrast to the reflectance spectrum of the much larger Pluto, 
which is rich in absorption
features due to CH$_4$, N$_2$, and CO \citep[see][for reviews]{1997plch.book..221C,2002AREPS..30..307B}, all
of which are volatile at the 30-50 K surface temperatures of Pluto. 
Triton, a body even larger than Pluto and
thought to be a captured KBO, has a similar compliment of 
volatile chemicals on its surface \citep{brown_triton}.
The advent of wide field surveys of the outer solar system 
\citep{2003EM&P...92...99T, 2008ssbn.book..335B, 
2010ApJ...720.1691S}, however, led to the discovery of objects approaching and 
even comparable in size to Pluto, many of which are also covered
in some of these same volatile chemicals. 

Definitive detections of CH$_4$ (a molecule with strong absorption
bands in the near-infrared which make the species easy to detect) have
now been made on 6 of the 8 largest known KBOs: Triton, Eris, Pluto,
Makemake, Sedna, and Quaoar 
\citep{brown_triton, 1997plch.book..221C, 2005ApJ...635L..97B,2006A&A...445L..35L, 2005A&A...439L...1B,2007ApJ...670L..49S}. A 7th -- 2007 OR10 -- 
is suspected from indirect evidence to also contain
CH$_4$, but no confirming spectrum has yet been obtained \citep{2011ApJ...738L..26B}.
N$_2$ and CO have weaker absorption bands and are therefore more
difficult to detect, so no absorptions due to these species have
been detected on any objects other than Pluto or Triton. 
Figure 1 
shows a composite of spectra of some of the volatile-containing
KBOs.
\begin{figure}
\includegraphics*{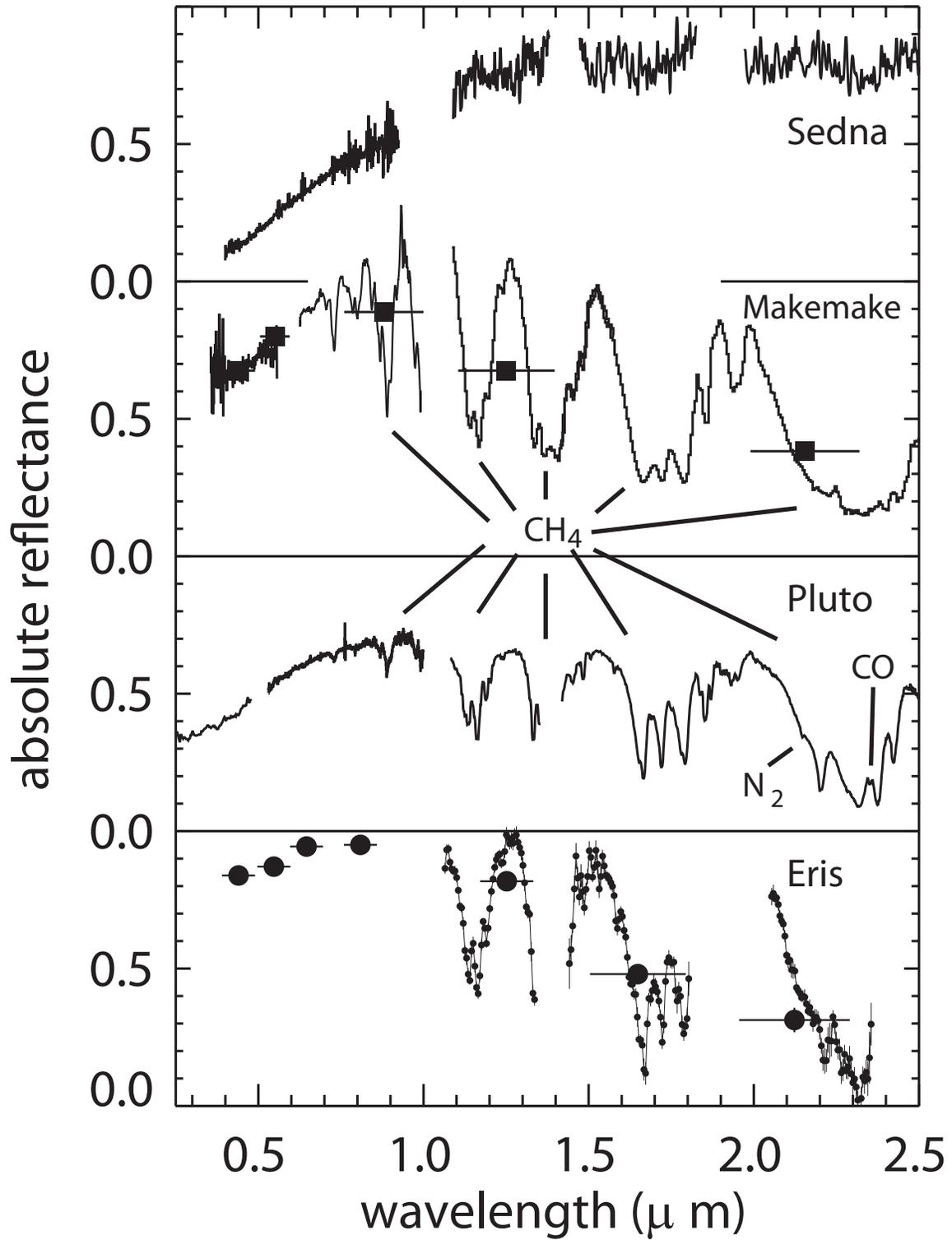}
\caption{
Spectra of four of the largest and most volatile rich KBOs.
The prominent CH$_4$ absorption lines can be seen
on all objects. The weaker CO and N$_2$ lines can only be detected
on Pluto.
}
\end{figure}

The specific volatile species detected on these largest KBOs are not 
surprising. Of the molecules know to be abundant in comets,
CH$_4$, N$_2$, and CO are the only ones that have moderate 
vapor pressures at the temperatures in the Kuiper belt. Most other
major molecules remain in the frozen state on the surface, while some
super-volatiles, like noble gasses, have such high vapor pressures that, if
they were present, they would be exclusively in the gas phase. 
CH$_4$, N$_2$, and CO, in contrast, can reach an equilibrium between
frost on the surface and a molecular atmosphere. This surface-atmosphere
exchange ensures that, when volatile abundances are sufficiently high,
the volatile frosts will coat the surface, masking the composition
of whatever lies beneath.

The relative abundances on the volatile containing KBOs are not identical.
On Makemake, the brightest and most easily studied of the newly
discovered large objects,  
 the CH$_4$ absorption features appear significantly
different from those on the other volatile-rich KBOs. The absorption
lines are broad and saturated, as occurs when the optical path lengths
through the solid CH$_4$ are long. The surface producing this
spectrum has been modeled as a solid slab of CH$_4$ with voids 
approximately $\sim$1 cm apart \citep{2007AJ....133..284B,2007JGRE..11206003E}. On Pluto and Triton, in contrast, 
small wavelength
shifts in the position of the 
near-infrared absorption features of the CH$_4$ show that the molecule
primarily occurs as a minor constituent diluted in an N$_2$
matrix \citep{1993Sci...261..742C, 1993Sci...261..745O}
though regions of pure CH$_4$ frost are also seen to exist \citep{1999Icar..142..421D}.
CH$_4$ is at least an
order of magnitude more abundant compared to N$_2$ than it is
on Pluto \citep{2007AJ....133..284B}.
The presence of N$_2$ has
nonetheless 
been indirectly inferred
by noting small shifts in the wavelengths of some of the far red
CH$_4$ absorption features \citep{2008Icar..195..844T}. 

Quaoar, even more dramatically, has such a low surface abundance of CH$_4$
that only the strongest absorption features are
detectable, superimposed on a background of water ice absorption
\citep{2007ApJ...670L..49S}. 
The CH$_4$ seems to be a patchy covering on an involatile water
ice substrate. The more recently discovered 2007 OR10 is thought to 
have a similar surface composition to Quaoar, based on its size and
its optical colors (see below). 

The presence of volatiles on these large objects and even the differences
in relative abundances between the volatiles can be explained by
a simple model of volatile loss and retention in the Kuiper belt
developed by \citet{2007ApJ...659L..61S}. In this model all KBOs are assumed to
start with typical cometary abundances but then volatile escape to space
occurs at a rate determined by the 
surface temperature and gravity. Volatile loss
can occur through a variety of mechanisms on a small body with 
a vapor-pressure controlled atmosphere, but the slowest mechanism
possible is simple Jeans escape from an atmosphere in hydrostatic
equilibrium at its surface equilibrium temperature. An updated
version of the model results 
of \citet{2007ApJ...659L..61S} (from \citet{2011ApJ...738L..26B}) is shown in Figure 2. In short, volatiles
\begin{figure}
\includegraphics*{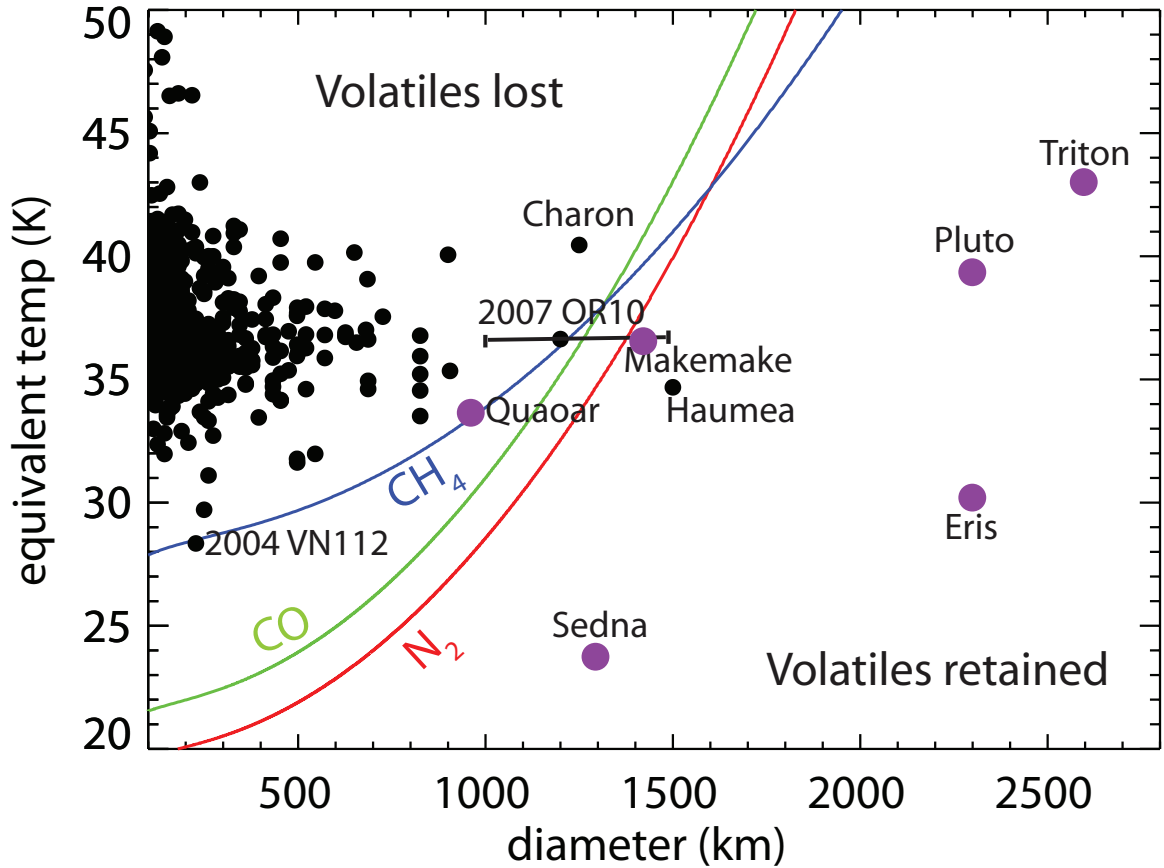}
\caption{Volatile retention and loss in the Kuiper belt.
Objects to the left of the CH$_4$, CO, and
N$_2$ lines are too small and too hot to retain any of those
surface volatiles
over the age of the solar system, while objects to the right can retain those
volatiles. All objects shown in purple have had CH$_4$ measured on their
surfaces. Some have additionally had N$_2$ or CO detected. No objects to 
the left of the lines have had any of these volatiles detected. Methane
is suspected, but not confirmed, on 2007 OR10.
}
\end{figure}
can be retained on objects which are either massive enough to prevent
significant Jeans escape or cold enough to prevent significant vapor
pressure of the frosts. The three main volatiles behave slightly
differently because of differing vapor pressures and molecular masses. 
N$_2$ or CO have identical molecular masses, so escape identically
for identical vapor pressures, but N$_2$ is more volatile, so N$_2$ always
has a higher vapor pressure and escapes more quickly. CH$_4$ has the 
lowest vapor pressure of the three, so generally escapes more slowly,
but for high enough temperatures the vapor pressure is sufficiently high
that the escape is controlled more by molecular weight, and the lighter
CH$_4$ escapes the quickest.

In the model, four important regimes are seen, corresponding nearly 
precisely to the observations. The largest and coldest objects --
Triton, Eris, Pluto, and Sedna -- have the ability to maintain
their full complement of volatiles. All three volatiles have
indeed been detected on Triton and Pluto. On Eris and Sedna
direct detection of N$_2$ is difficult owing to the faintness of
the objects, the weakness of the single quadrapole N$_2$ absorption
line, and the fact that at the low temperatures of the very distant
Eris and Sedna, N$_2$ should be in its $\alpha$ state leading to
an extremely narrow absorption feature \citep{1995Icar..116..409T}. 
On Eris, however,
subtle shifts in the far red portion of the spectrum again show
evidence that some CH$_4$ is diluted within a N$_2$ matrix
\citep{2010ApJ...725.1296T}. 
For Sedna, early observations suggested the 
presence of an N$_2$ absorption feature, but this feature appeared
similar to that on Triton, a broad feature of $\beta$-nitrogen, which
would be difficult to explain on the frigid Sedna \citep{2005A&A...439L...1B}. 
Other observations
have failed to confirm this feature \citep{2010AJ....140.2095B}. It
appears likely that N$_2$, even if abundant on Sedna, would be 
nearly impossible to detect spectroscopically. CO would be extremely
difficult to detect on Eris or Sedna, and, to date, no meaningful upper
limits have been placed.

The next major regime in the model is occupied only by Makemake. Makemake
has just the right size and temperature to be able to retain its CH$_4$,
but to be on the edge of not being able to retain N$_2$ and CO. The
unusual surface of Makemake appears well explained by this model. 
With a low abundance of N$_2$ on the surface, CH$_4$ becomes the major
constituent and anneals into large slabs, giving rise to the long optical
path lengths which saturate the infrared spectrum. 

The third regime in the model is the transition where CH$_4$ can only
barely be retained and other volatiles are nonexistent. 
Quaoar and potentially
2007 OR10 are both in 
this regime. These objects are both sufficiently depleted even in methane
that their water ice substrate is visible in regions not covered by 
methane. 

The object 2004 VN112, though likely small, has a perihelion of 47 AU and thus
stays cold enough to potentially retain at least CH$_4$. 
This object is too faint for infrared 
spectroscopy, so we have no direct
indication of its surface composition, and its albedo, and thus, size
is unknown. Its optical colors are relatively
neutral, however, suggesting that it is not dominated by CH$_4$ irradiation
products (see below). Further physical study of this object is clearly warranted.

The final regime within this model is the region in which most of the
Kuiper belt resides, where temperatures are too high and 
masses are too low, so that even with slow Jeans escape, the three main
volatiles must be depleted over solar system time scales.

To date, this model has been flawless for determining which
objects do and do not retain volatiles in the Kuiper belt (with 
the important exception of Haumea, which is discussed 
below). It is interesting -- and even unexpected -- that 
such a simple model would work so well. The well-studied 
atmospheres of
Pluto and Triton, for example, do not conform to the simple assumptions
of surface temperature hydrostatic equilibrium and Jeans escape. Perhaps,
however, the power of the model is that as volatiles begin to become
depleted through faster mechanisms such as hydrodynamic escape, the atmosphere
eventually becomes tenuous enough that Jeans escape is the only remaining
mechanism
functioning. Jeans escape is then so slow compared to other processes
that it eventually
dominates the total amount of time that it takes for volatile loss.

While Kuiper belt volatiles have mostly been studied from infrared
spectroscopy, stellar occultations provide another alternative for
study of volatile compositions. 
Such occultations should be able to
provide insights into N$_2$ abundances on Eris, Makemake and Sedna
and occultations should provide the best means of looking for
volatiles around small distant objects such as 2004 VN112.

\subsection{Radiation processing}

All bare surfaces in the solar system are subject to irradiation by
solar wind, UV photons, and cosmic rays, all of which are 
capable of inducing chemical changes in the surfaces. The surface
colors of most KBOs are likely set by this irradiation (see below), but,
for many years, the only radiation products thought to be seen
in the outer solar system were
tholins --  the reddish involatile 
residues of long term irradiation 
of hydrocarbon and nitrogen containing compounds.
 Many other
simpler molecular
radiation products of N$_2$, CO, and CH$_4$ were synthesized
in laboratories and predicted to be present on Pluto or Triton
\citep{2008ssbn.book..507H}, but the low abundances, difficult-to-observe spectral
features, and lack of adequate data prevented any positive detections.

The discovery of the large slabs of CH$_4$ on Makemake, however, 
also led to the first direct detection of simple
radiation products. The saturation of the strong
CH$_4$ absorption 
line around 2.3$\mu$m allows deviations from the CH$_4$ spectrum
to be easily identified. The strongest deviations are clearly due to the
presence of C$_2$H$_6$ (Fig 3). C$_2$H$_6$ forms
%figure 3
\begin{figure}
\plotone{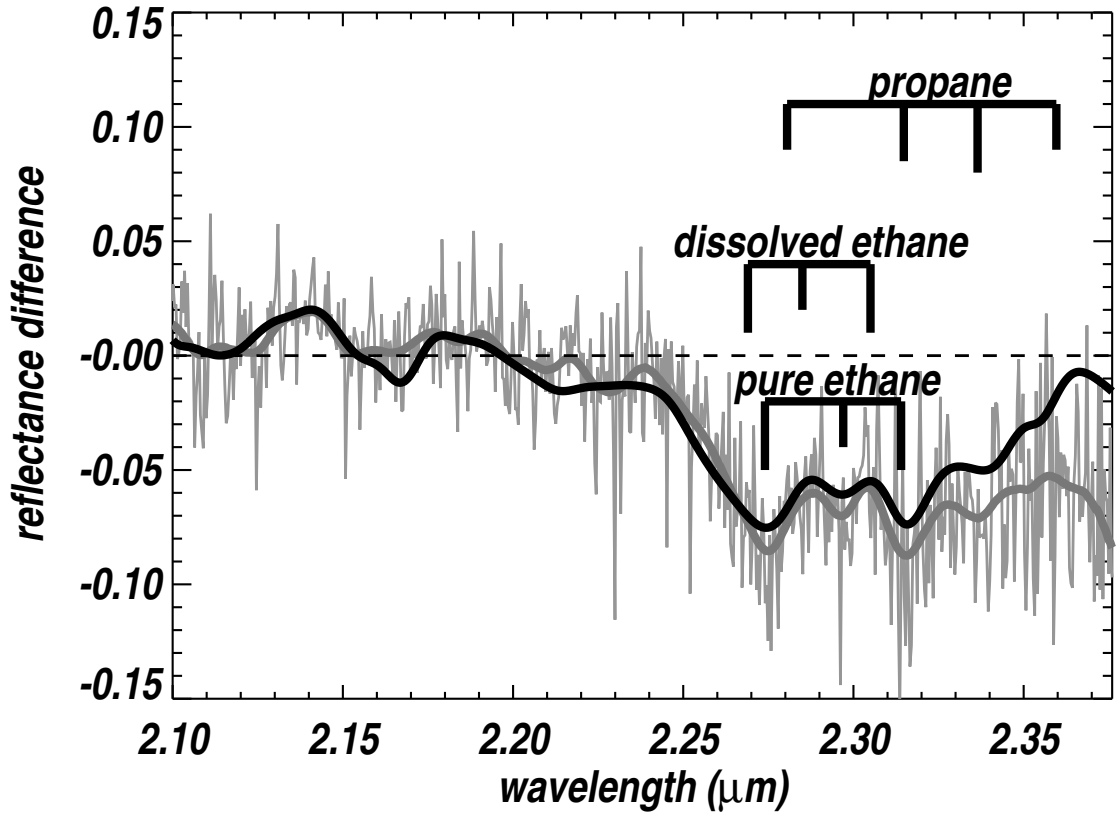}
\caption{
Detection of pure ethane on Makemake. The data show the difference between
the data and a CH$_4$ model of the surface. The deviations from the CH$_4$
model show the clear signature of pure ethane ice, an expected 
irradiation product of CH$_4$. Evidence for additional absorption
is present, but no positive identification of the species has yet
occurred.}
\end{figure}
from the combination of two CH$_4$ molecules which have each had a 
hydrogen atom removed by irradiation, and it is the first stable molecule
to form in the presence of CH$_4$ irradiation \citep{2006ApJ...653..792B}. 

Makemake, which is dominated by CH$_4$ rather than N$_2$, as on Pluto,
is an ideal laboratory for understanding this radiation processing.
On Pluto, creation of C$_2$H$_6$ is stymied by the dilution of the
CH$_4$ in a N$_2$ matrix. Radiation can remove a hydrogen atom, 
but another CH$_3$ radical will rarely be present to form C$_2$H$_6$.
Nonetheless, recent observations have suggested a small amount
of C$_2$H$_6$ on Pluto, as well \citep{2010Icar..208..412D,2010Icar..210..930M}, presumably coming 
from the small regions of pure CH$_4$ which have been detected
(see above). 

Quaoar, though having significantly less coverage of CH$_4$ on its
surface than Makemake, also appears to have detectable C$_2$H$_6$ 
embedded within its surface, which would be expected for an object
with CH$_4$ and little to no N$_2$. Quaoar is also unusual for
having both deep water ice absorption features and a very 
red color. The radiation processing of CH$_4$ appears
likely to be the cause of this coloration; long-term irradiation
of CH$_4$ will cause the ice to turn red \citep{2006ApJ...644..646B}.
2007 OR10, which appears in a similar regime
of volatile retention as Quaoar, has a similar very red surface
combined with deep water ice absorptions, leading to the hypothesis
that it, too, contains CH$_4$, and, presumably, its radiation 
products.

After the creation of C$_2$H$_6$, continued irradiation leads to
C$_2$H$_2$ (acetylene) and C$_2$H$_4$ (ethylene).
These molecules have increasingly difficult to detect spectral
features and have not been detected on Makemake -- or any other
object in the solar system -- to date, but it is likely that they
are present as temporary products which eventually lead to
the red coloration of the long-chain hydrocarbons.

Orcus is similar in size to Quaoar but warmer, and thus is not
expected to retain volatiles. \citet{2010A&A...520A..40D} nonetheless suggest
the detection of C$_2$H$_6$, an involatile irradiation 
product on this surface. 
While it is possible to imagine a scenario in which CH$_4$ irradiation
leads to irradiation products which then stay behind as a lag deposit
after the CH$_4$ has been lost, it is difficult to construct a scenario
in which the surface does not also retain the red coloration of
more complex irradiation products. Orcus is instead amongst the most
neutrally colored KBOs. We suspect the higher signal-to-noise spectra
of Orcus will confirm a lack of methane irradiation products on its surface.

\subsection{Haumea and its family}
Haumea is unique in the solar system. It is the fastest rotating and
most elongated
gravitationally bound object in the solar system, has a density
nearly that of rock \citep{2006ApJ...639.1238R}, and is surrounded by two
moons \citep{2005ApJ...632L..45B, 2006ApJ...639L..43B}. Though it is one of the largest
objects in the Kuiper belt and would be predicted by the simple
model above to contain abundant volatiles, its surface is instead
dominated by what appears to be nearly pure water ice \citep{2007ApJ...655.1172T}.
The moons likewise appear to have surfaces of 
pure water ice \citep{2006ApJ...640L..87B, 2009ApJ...695L...1F},
and a dynamical family of KBOs with pure water ice surfaces
exists in orbits separated from Haumea by only a few hundred meters per
second (Fig 4).
\begin{figure}
\includegraphics[scale=.75]{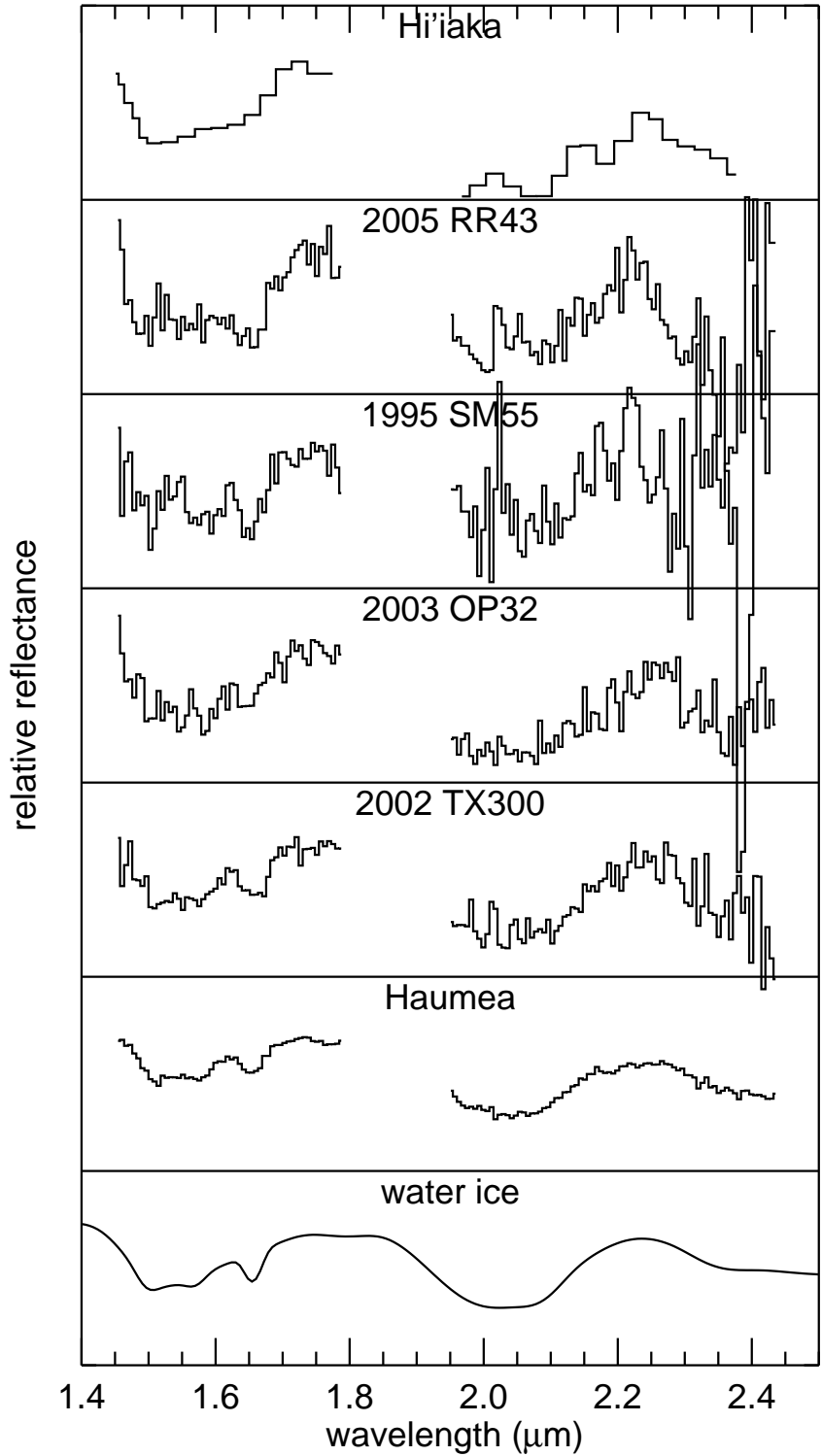}
\caption{
Near infrared reflectance spectra of Haumea, its satellite Hi'iaka, 
and some members of its collisional
family, compared to a model of a laboratory spectrum of pure ice.
No objects with comparably deep water ice absorption features are
found anywhere in the Kuiper belt other than the dynamical
vicinity of Haumea. The spectra are consistent with nearly pure
water ice. In all cases with sufficient signal-to-noise, the
spectrum shows the 1.65 $\mu$m absorption feature of crystalline ice.
This feature is ubiquitous in the outer solar system on objects
large and small and appears to be caused by exogenic rather than
endogenic processes.}
\end{figure}

While Haumea is large enough to be differentiated, so that we can imagine
an icy outer layer surrounding a predominantly rocky body, the
known dynamical family members and the satellites range in size from $\sim$70
to $\sim$ 370 km (assuming 0.7 albedos), likely too small to be differentiated. For these objects,
the fact that the surfaces appear to be nearly pure water ice
strongly suggests that these objects are nearly pure water ice
in their interiors, also. In particular, the mutual orbital
perturbations of the two Haumea satellites show that their masses are
consistent with them having bulk densities even lower than 1 g cm$^{-3}$
\citep{2009AJ....137.4766R}.

All of these characteristics are likely the product of a single
oblique giant impact onto a differentiated proto-Haumea earlier 
in the history of the solar system \citep{2007Natur.446..294B}. Prior to impact, the proto-Haumea
would have been differentiated into a rocky core, a relatively pure water
ice mantle, and a crust of volatiles that had been driven to the surface
and their irradiation products.
The oblique impact
then gave Haumea its fast rotation, causing its
extreme elongation. In addition, the impact must have blasted off
a significant fraction of the proto-Haumea's icy mantle, which then
became the satellites and the collisional family.

In this scenario, the nearly pure water ice surface of Haumea (along with
its high density) occurs because we are viewing the exposed layers of the
interior pure icy mantle. The nearly pure water ice surfaces of
the dynamical family members and the satellites occur because they are
fragments of the interior of the mantle. Long term irradiation
of these pure ice fragments does not cause any of the reddening or
darkening that would be expected if any hydrocarbons were present.
 Fragments of the crust, which
should have contained more irradiated hydrocarbons, presumably were
created also, but the volume of crust is much smaller than that of
mantle, and there is currently no method for
positive identification of non-ice components of
Haumea, so to date only mantle fragments are known.

Even if the post-impact
Haumea and its collisional family originally began with
nearly pure water ice surfaces, the fact that these surfaces
remain relatively uncontaminated with dust, fragments of dark
impactors or irradiated hydrocarbons is a surprise. It has even
been suggested that the collision must have been relatively recent
for the surfaces to appear so pure \citep{2008AJ....136.1502R}. Dynamically 
modeling of the diffusion of Haumea with the 12:7 mean motion resonance
with Neptune suggests, however, that the collision occurred billions of
years ago \citep{2007AJ....134.2160R}. Modeling of collision probabilities suggests
that the collision must have occurred near the beginning of the solar 
system \citep{2008AJ....136.1079L}. Pure water ice surfaces in the Kuiper must be
capable of staying relative pristine for billions of years. While
the reasons for this longevity are not clear, we will use this empirically
determined fact below when discussing water ice on the mid-sized objects.

In addition to being relatively pristine, the spectrum of water ice
on Haumea and its family members always contains the 1.65 $\mu$m
absorption feature due to crystalline water ice. While this 1.65 $\mu$m
feature appear ubiquitous throughout the outer solar system, its
presence is perhaps surprising. Irradiation studies suggest that
crystalline water ice should turn to amorphous water ice on
relatively short time scales unless some heating is applied to
recrystallize the water ice. This 1.65 $\mu$m feature is thus
often taken as evidence of some type of internal heating, from
cryovolcanism \citep{2004Natur.432..731J} 
to radioactivity and tidal forces \citep{2011A&A...528A.105D}. The fact
that the 1.65 $\mu$m absorption can be seen 
even on Haumea family members as small as 1995 SM55, with
an estimated size of $\sim$180 km in diameter (assuming
an identical albedo to Haumea) 
demonstrates, however, that no
such internal mechanism is required for the appearance of this feature.
These objects are far to small to maintain the liquid water beneath
their surface that would be necessary to support any current
cryovolcanism  and as fragments of the icy mantle of 
Haumea they are thoroughly lacking in the rocks that would give rise
to radioactive heating. It seems clear, even if not understood, that 
the appearance of the 1.65 $\mu$m absorption feature of crystalline
water ice does not contain any information about internal processes,
but rather contains new information about the physics of crystallization
under these conditions.

\section{Mid-sized objects}
While the surface compositions of the largest object appear well 
understood and even predictable, less attention has been paid to the
mid-size $\sim$500 - 1000 km diameter KBOs which are just a little
too small to be able to have retained surface volatiles. In some
ways, these objects form a more interesting class than the 
larger objects, as these objects do not have a frosty veneer
hiding the inherent surface composition. Unfortunately, however, while
high quality spectroscopy of the largest and brightest objects
can be readily obtained, these mid-sized objects are considerably more
difficult to observe even with the largest telescopes in the world. 
Major observing campaigns with the Keck telescope \citep{2008AJ....135...55B,
specsub} and
the VLT \citep{2009Icar..201..272G,2011Icar..214..297B} have provided much of what we know of
the surface compositions of these objects.
\subsection{Water ice}
\citet{2011Icar..214..297B} and \citet{specsub} both analyze extensive
collections of high quality spectra of KBOs and their progeny.
In \citet{2011Icar..214..297B} the possibility of water ice in VLT spectra
is first assessed by
simply calculating the depth of a potential 2 $\mu$m absorption feature
and then followed by detailed modeling including many different components.
\citet{specsub} concentrate on a examining the possibility of water
ice in Keck spectra by fitting a simple water ice plus sloped continuum
model to the near infrared spectrum. 
The VLT and Keck results are in agreement both on the broad
results and on most of the individual objects. 
\begin{figure}
\plotone{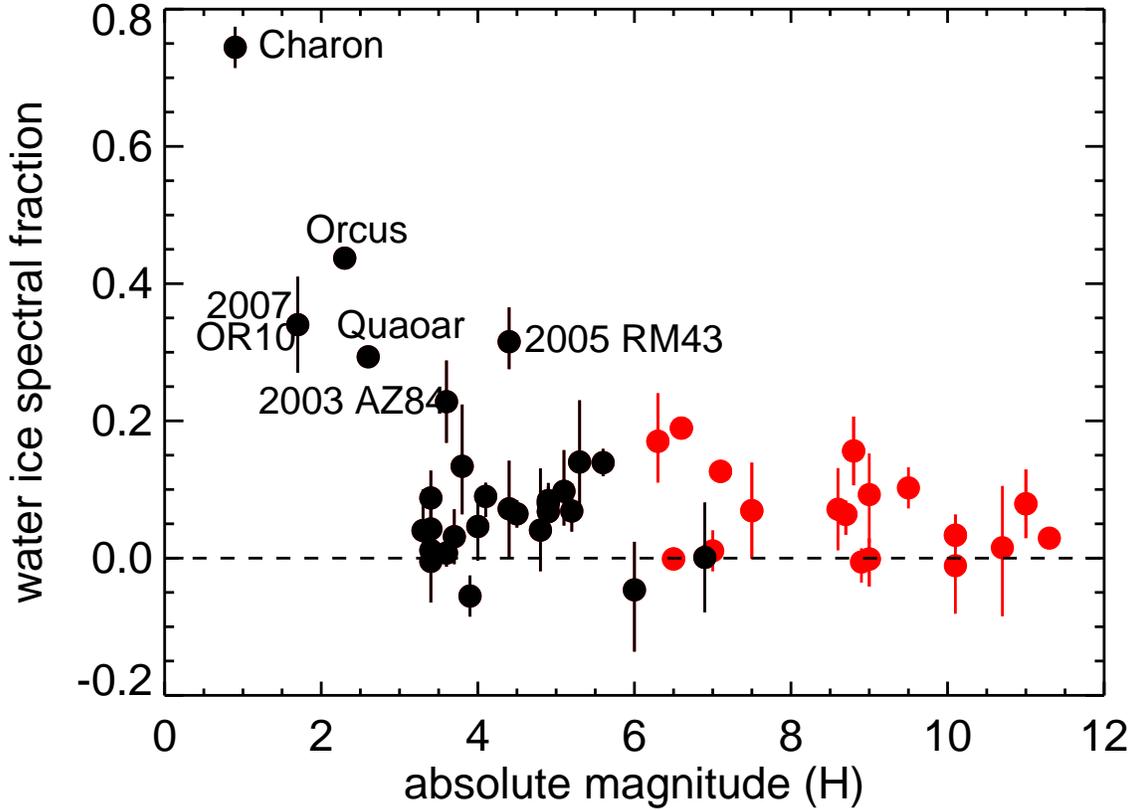}
\caption{
The water ice spectra fraction -- a measure of the amount of
water ice absorption in an object spectrum -- as a function of
absolute magnitude. Absolute magnitude is a measure of the intrinsic brightness
of the object, which is related to size. The largest objects -- on the left of the plot --
have the smallest absolute magnitudes. A clear trend is seen for the largest
objects to have the deepest water ice absorption spectra. The
transition around $H\sim4$ from moderate to deep water ice
absorption is a potential indicator of the size at which
interior oceans caused surface ice flows in the past.
The water ice fraction on smaller objects is indistinguishable
from that on the even smaller centaurs.}
\end{figure}

In Figure 5, we show the water ice spectral fraction -- a measure
of the depth of the water ice absorption in the spectrum -- as 
a function of the absolute magnitude (a proxy for diameter) 
of KBOs and also of centaurs (see below),
from the \citet{specsub} sample.
A clear trend is evident:
brighter than an absolute magnitude of $H=3$, all KBOs have deep
water ice absorption (water ice spectral fraction $>0.2$). Fainter than an absolute magnitude of $H=5$, deep water ice absorption is never seen.

While such a correlation of absolute magnitude and presence of
water ice might be expected simply from the increased albedo
of objects with more water ice on their surfaces, Spitzer radiometry
has shown that the $H>3$ KBOs are indeed smaller than the $H<3$ KBOs
\citep{2008ssbn.book..161S}.
Somewhere between the $\sim$650 km diameter of Ixion and 
the $\sim$900 km diameters
of Quaoar and Orcus, KBOs appear
to dramatically change in surface composition.

The change in surface composition is clearly not monotonic with
size. Ixion, 2002 UX25, 2002 AW197, 2004 GV9,
and  2003 AZ84 have $\sim$650 km diameters
within their uncertainties \citep{2008ssbn.book..161S}, but
only 2003 AZ84 has a large abundance of water ice on the surface. 
Likewise, the smaller 2005 RM43, which would have a diameter of 
$\sim$520 km if it has the same albedo as 2003 AZ84, has water
ice absorption as deep as the larger objects Quaoar and 2007 OR10,
while the similarly-sized objects Varuna and Huya have much
smaller absorption depths. 

Smaller than the $\sim$520 km size of 2005 RM43, however, no (non-Haumea
family member) KBO has been found with strong water ice absorption.
Few objects in this range are bright enough for
high quality spectroscopy (1996 TP66, with a diameter of $\sim$180 km
for an assumed 0.1 albedo is the smallest KBO with a well measured
spectrum), however. Nonetheless, spectroscopic evidence from
centaurs and photometric evidence from smaller KBOs (see below)
suggests that this trend continues, and that between 500 and 700 km
in diameter KBOs transition from typical surfaces of small KBOs
to those dominated by absorption due to water ice.

Additional evidence for a surface change in this diameter
range comes from measurements
of albedos. 
Though the size and
albedo data available from Spitzer photometry have large
uncertainties, a jump in albedo at this size range 
is apparent.
The same effect is see in the
cumulative absolute magnitude distribution of the large
objects. While this distribution follows a power law from 
absolute magnitudes of 3 until at least 4.5, for objects 
brighter than $H=3$, the cumulative number distribution is
larger than would be expected from this power law, a clear
sign of the increased albedos of these mid-sized
objects \citep{2008ssbn.book..335B}. Implications for this surface
change are discussed below.

\subsection{Ammonia}
Of these mid-sized objects with strong water ice absorption
features, two have evidence for a 2.25 $\mu$m absorption
feature due to ammonia. Ammonia was first detected on Charon
\citep{2000Sci...287..107B,2007ApJ...663.1406C,2010Icar..210..930M} and later suggested to be present also on
Orcus \citep{2008A&A...479L..13B}. In both cases, the presence of ammonia
was postulated to be due to the flow of ammonia-rich interior
liquid water on to the surface of the object at some point in
the past. Detailed models of the interior structure and evolution
of these two bodies have suggested that such a scenario 
appears reasonable \citep{2007ApJ...663.1406C, 2010A&A...520A..40D}.

Ammonia has not been detected on the other water ice-rich objects,
but for Quaoar a bright object where high quality
spectra can be obtained, the 2.3 $\mu$m region of the spectrum
where ammonia has its only detectable near-infrared feature is
instead dominated by the strongest absorption feature due to CH$_4$.
Ammonia has additional absorption features beyond 3 $\mu$m which
could be distinguished from CH$_4$ but spectral measurement
at these wavelengths awaits larger ground or space-based telescopes.
The smaller water-ice-rich objects 2003 AZ84 and 2005 RM43 have spectra with
insufficient signal-to-noise to detect ammonia. Larger telescopes will
again be required.

Based on the likely presence of ammonia on Charon and Orcus,
and the sharp increase in water ice absorption with larger
size for these objects, 
we hypothesize that on these largest objects the presence
of water ice -- whether ammonia is detectable or not -- is a
remnant of past liquid flows on the surface. We predict that
for these objects ammonia will always be detected when sufficient
signal-to-noise is available. The liquid flows
need not be recent -- the purity of the water ice on the surface of
Haumea and its family shows that water ice can remain pristine through
the age of the solar system -- but the liquid flows must have 
occurred after the volatiles whose irradiation would cause coloration
(see below) have all escaped. The increase in water ice absorption
with size would be a natural consequence of the larger 
interior liquid reservoirs of larger objects. The sharp increase
in water ice absorption starting at diameters around $\sim$650 km
gives an important clue into the physics of liquid interiors
and surface water flows.

\section{Small objects: spectroscopic constraints}
\subsection{Kuiper belt objects}

Many objects smaller than the size where strong water ice
absorption is present are statistically
consistent with having no water ice on their surface, but there is
a significant prevalence in the data for positive detection of water ice.
Random and known systematic errors would not produce a bias towards
water ice detection. Indeed, we regard the nearly complete lack of 
objects with negative water ice fraction as an indication of the
robustness of our method. We conclude, therefore, that even the low
level of water ice fraction detected in the majority of the
objects is a real indication that water ice at a low level
is common even on the objects smaller than Ixion.

2002 VE95, the smallest Kuiper belt object with a very high quality 
spectrum ($\sim$330 km diameter assuming a 0.1 albedo), shows
clear evidence of crystalline water ice on its surface
\citep{2008AJ....135...55B, 2006A&A...455..725B}.
Indeed, the 1.65 $\mu$m
absorption feature of crystalline water ice can always be detected when
the signal-to-noise is sufficiently high. 

Based on their smaller sizes, it
seems likely that the water ice on these smaller objects is
not caused by liquid flows on the exterior, but rather by the
exposure of crustal water ice. An important test of this expectation
would be that these smaller KBOs should not show the presence
of ammonia on their surfaces. Currently feasible spectroscopy
cannot achieve the signal-to-noise required to make this test, however.

\subsection{Centaurs}
KBOs smaller than those discussed above are too faint for high quality
spectroscopy even with the largest telescopes in the world. To understand
the compositions of these objects, we have to resort to
observational proxies. One important
proxy has been spectroscopic observations of Centaurs. Centaurs are
former KBOs which have been perturbed onto short-lived planet-crossing
orbits. Being much closer to the sun than typical KBOs, they are 
brighter and easier to study in detail.

Spectroscopically, the Centaurs appear indistinguishable
from the smallest KBOs whose spectra can be measured: objects contain
either no detectable absorption features, a small amount of absorption 
due to water ice, or (in one case) absorption due to water ice and 
methanol. Figure 5 includes water ice absorption depth for a large
sample of Centaurs, compared to KBOs. No discernible difference
can be seen between the largest of the centaurs and the smallest of
the measured KBOs. 
The amount of water ice absorption seen on the surface of a centaur
does not appear to correlate with anything, including
perihelion, semimajor axis, optical color, dynamical lifetime, or 
activity. Interestingly, while much speculation has occurred about 
Centaur surface evolution as objects move closer to the sun and 
begin heating, infrared spectroscopy shows no such evidence of any
change in the distribution of water ice absorption depth.

\subsection{Methanol}
An absorption band at $\sim$2.27 $\mu$m
was first detected on the bright
centaur Pholus \citep{1993Icar..102..166D}, and \citet{1998Icar..135..389C} present 
the case that this band is plausibly due to the presence of
methanol, though they point out that the identification is
not unique and that other low molecular weight hydrocarbons
or photolytic products of methanol might fit the spectrum equally well.
Pholus is one of the reddest objects in the solar system, again
fitting our picture of optical colors of irradiated hydrocarbons well.

Based on lower signal-to-noise spectra with properties
which resemble the water-ice-plus-methanol spectrum of Pholus, the presence of
methanol was suggested on the KBOs 1996 GQ21 and 2002 VE95
\citep{2008ssbn.book..143B}. 
To examine this possibility more closely, we combine the Keck spectra
of these two objects to increase signal-to-noise and consider
the presence of methanol. While the signal-to-noise
remains low, the presence of absorption features similar
to those on Pholus is certainly plausible.
Both objects, like Pholus, are red.

A handful of other KBOs have recently been reported to also have
absorption features near 2.27 $\mu$m, but, unlike
Pholus, 1996 GQ21 and 2002 VE95, to not have absorption
due to water ice \citep{2011Icar..214..297B}. The signal-to-noise in
the spectral region are low,
so it is difficult to determine
if the absorption features are real. 
To examine the possibility that
a $\sim$2.27 $\mu$m absorption feature occurs on faint KBOs,
we first examine all of the KBOs and centaurs in the Keck sample
(which includes none of the potential methanol objects from
the VLT sample).
We find that a small number of objects have absorption features
at or near the 2.27 $\mu$m methanol absorption line, but
no single spectrum is sufficiently reliable in this region to
assert a positive detection. To increase the signal-to-noise,
we sum the spectra of all of the KBOs and centaurs in the Keck
sample
except for those with water ice absorption at the level of
that seen on 2003 AZ84 or deeper and those already suspected
to contain methanol-like features (1996 GQ21 and 2002 VE95). This combined spectrum shows
residual absorptions due to water, as would be inferred from
the positive detections on most objects. The only major deviation
from the water ice spectrum occurs at precisely the wavelength
of the feature seen on Pholus and suspected on 1996 GQ21 and 2002 VE95.
We conclude that
the methanol-like feature is indeed present at a low-level on
KBOs even though the feature cannot be reliably identified in
individual spectra
(Figure 6).
\begin{figure}
\includegraphics[scale=.75]{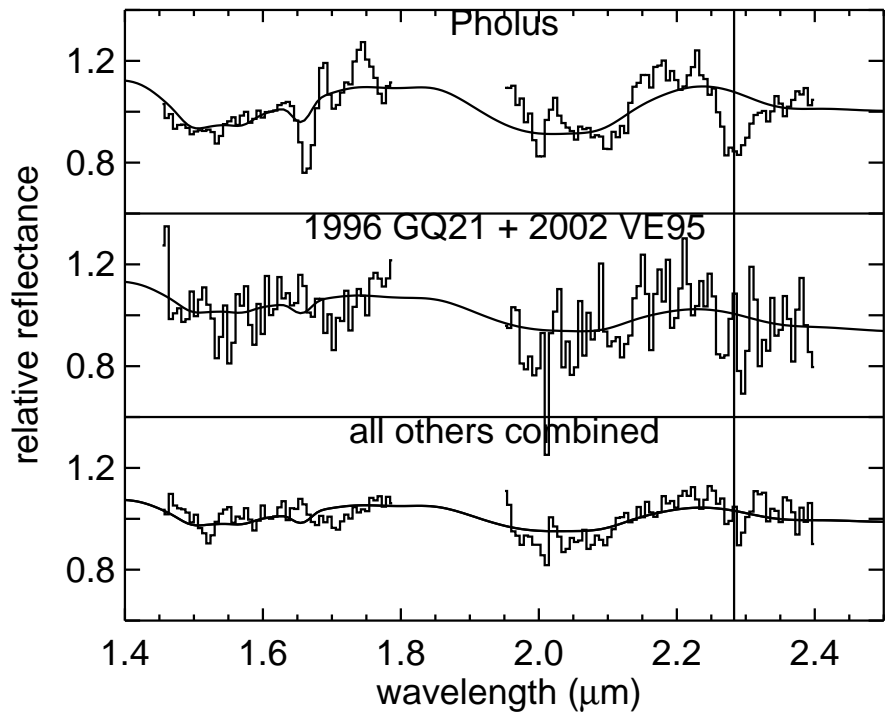}
\caption{The region of the near-infrared spectrum containing 
the 2.27 $\mu$m absorption feature attributed to methanol.
The feature can clearly be seen on Pholus, superimposed on a 
small amount of water ice absorption. The combined
(to increase signal-to-noise) spectrum of 1996 GQ21 and 2002 VE95
shows hints of a feature at the same location. The sum of 38
KBO and centaur spectra, none of which individually show clear evidence
for the feature, clearly shows a feature at the same spot.
}
\end{figure}

No hypothesis has ever been formulated for the 
sporadic presence of methanol on KBOs or centaurs, other
than to point out that methanol is common in cometary comae,
so it is expected to be present in the interior of KBOs.
Its presence is less expected on the surface of KBOs, however,
as the absorption features of hydrocarbon ices quickly degrade
under irradiation while the remnants turn red \citep{2006ApJ...644..646B}.
Visible methanol absorption features suggest that the
methanol has only recently been exposed at the surface, perhaps
as a result of a collision exposing the subsurface. 
One prediction of this suggestion would be that the amount of
methanol observed would vary as different faces of the object were
observed. While such a test is possible in principle, in practice
spectroscopy of these faint objects 
is sufficiently difficult that variation would be difficult to
prove. If, however, irradiation preferentially turned the regions
with exposed methanol red, these objects should perhaps show color
variation with rotation, something which is otherwise rarely 
observed in objects of this size \citep{2008ssbn.book..129S}.

\subsection{Silicates}
In addition to ices and their irradiation products, we should
expect that KBOs that are small enough to be
undifferentiated should expose some of their
rock component on the surface. While silicates such as
olivine or pyroxene
are often included in detailed models of KBO spectra 
\citep[i.e.][]{2010Icar..208..945M,2011Icar..214..297B}
 no
specific absorptions are easily observable: olivine, for example, has a 
broad absorption centered at around 1 $\mu$m, where observations
are usually poor. These silicates are thus included in models
to fit the overall spectral shapes over the $\sim$0.6 - 1.2 $\mu$m
range.

Mid-infrared observations have the possibility of positively
identifying silicate emission in the 10 and 20 $\mu$m region 
on KBOs. To date only the centaur Asbolus has had a 
spectrum measured with sufficient signal-to-noise to even
detect these spectral regions, and, in this case, emissivity
peaks around 10 and 20 $\mu$m are indeed seen, similar to
that observed in Trojan asteroids \citep{2006Icar..182..496E,2008ssbn.book..143B},
and interpreted to be due to fine-grained silicates.
Future observations will require improved space-based 
mid-infrared facilities, but characterization of 
actual silicate composition is indeed possible.

While olivines or pyroxenes cannot be specifically identified
in the visible to near-infrared range,
there have been a few reports of shallow broad absorption
features in the visible wavelength range
similar to absorptions seen on some asteroids.
On asteroids these are generally attributed to aqueously 
altered silicates.
Confirmation of these features has been difficult; the absorptions
are subtle and have frequently appeared changed or absent upon reobservation
of the same object \citep{2004A&A...421..353F,2004A&A...416..791D, 2009A&A...508..457F, 2008A&A...487..741A, 2003AJ....125.1554L}
While these changes are usually attributed to rotational
variability, it is worth noting that this speculation has never
been verified. Indeed, for one object observations over half 
of a rotational period showed no signs of the visible absorption 
features.

While it is possible that these difficult-to-confirm features 
are a product of sporadic systematic error, the possibility
of the existence of aqueously altered silicates is 
an interesting one to consider. The presence of
liquid water on surfaces in the Kuiper belt might seem surprising, but
hydrous materials seem to be present in small comets, interplanetary
dust particles, and debris disks \citep{2004A&A...416..791D}. 
If these detections are indeed real, the most surprising thing
about them, perhaps, is that they are uncommon.

%%%%%%%%%%%%%%%%%%%%%%%%%%%%%%%%%%%%%%%%%%%%% switch to PHOTOMETRY

\section{Small objects: photometric constraints}
Small objects in the Kuiper belt are too faint to detect spectroscopically,
but photometric measurements can still give information -- albeit limited --
about the surfaces of these objects. To date, the single most robust
conclusion based on photometry is that Kuiper belt surfaces are
diverse.
Large surveys of KBO optical
colors have found a wider range of
surface colors
in the Kuiper belt than in any other small body population in the 
solar system. The colors are uncorrelated with
most dynamical or physical properties \citep[see review in][]{2008ssbn.book...91D}.
A few systematic patterns
have been found, however, which are important clues to understanding the
surface compositions of these outer solar system objects.

\subsection{The bifurcated colors of centaurs} 
The range of centaur optical colors generally covers the same wide
 range of colors
found in the Kuiper belt, but the centaurs are deficient in colors in
the middle part of the range, giving the distribution of centaur optical
colors
a bimodal appearance \citep{2008ssbn.book..105T} with a neutral
and a red clump of objects. Interestingly, this bifurcation is not
seen only in the centaurs, but it appears to extend to all objects with
low perihelion distance whether the objects are dynamically unstable or
not. This result immediately suggests that the bifurcation in the 
optical colors is somehow formed through the increased heating or
irradiation experienced by lower perihelion objects.

The H/WTSOSS (Hubble/WFC3 Test of Surfaces in the Outer Solar System) survey which used the Hubble Space Telescope to extend
 optical photometry of centaurs (and other low
perihelion objects) into the near-infrared
\citep{hwtsoss} found that the neutral and red clumps do not consist
of two groups with identical surfaces, 
but rather are best described by two groups that fall along two
separate 
mixing lines.
The neutral clump of objects consists of a mixture of
a nearly neutrally reflecting material and a slightly red material,
while the red clump of objects consists of a mixture of the same
neutrally reflecting material and a much red material
(Fig 7). 
%figure hwtsoss
\begin{figure}
\includegraphics[scale=.65]{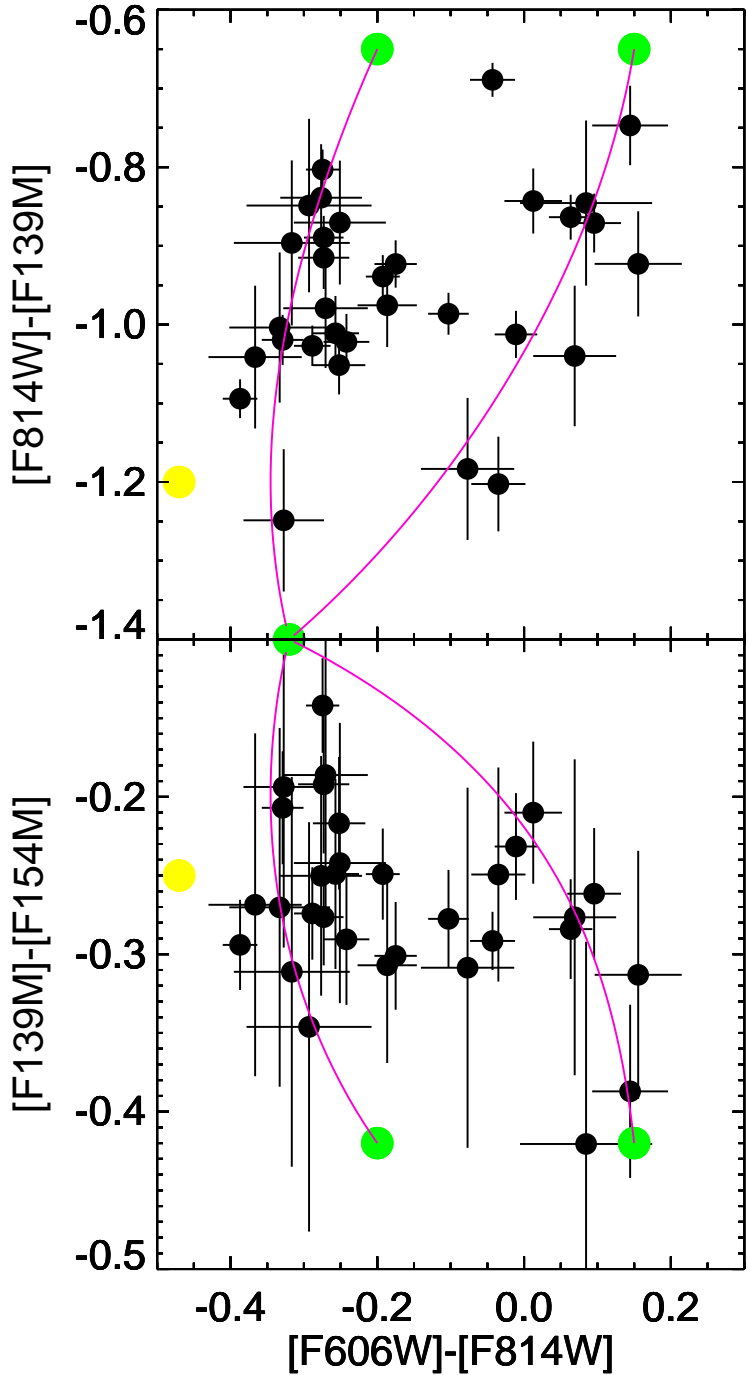}
\caption{
Three-color HST photometry
of objects with perihelia inside of 30 AU from
the H/WTSOSS survey. The filters are at 0.606, 0.814 1.39, and
1.54 $\mu$m, corresponding roughly to $R$, $I$, $H$-continuum, 
and a filter sampling the 1.6 $\mu$m water ice absorption 
feature. 
The yellow point shows the
colors of the sun. 
The objects appear to bifurcate into two clumps, one with near-solar optical colors
(the ''neutral'' clump) and one with significantly redder optical colors (the ''red clump'').
The neutral clump (on the left in [F606W]-[F814W]
colors) shows a clear spread in the other colors, while the
red clump shows a spread in all three colors. The color variation in
both clumps can be described by a mixing line (purple lines) where
the neutral and red clumps both have a single 
common end member but are mixed with either a more neutral or more 
red component, respectively.}

\end{figure}

While three color photometry is generally incapable of identifying specific 
ices or minerals, \citet{hwtsoss} find that the neutral component 
common to all centaurs is consistent with some of the same
hydrated silicates suggested from the optical spectroscopy
discussed above. Significantly more spectral work is required,
however, to further explore this possibility.

 \subsection{The uniform colors of the cold classical KBOS.}
While most of the rest of the 
Kuiper belt appears to be composed of essentially the
same distribution
 of neutral to red objects \citep{morbandbrown,2008ssbn.book...91D}
one dynamical region stands out for its
homogeneous composition. The cold classical Kuiper belt was first identified
as a dynamically unique region of the Kuiper belt -- a difficult-to-explain 
overabundance of low inclination, dynamically cold objects beyond about
41 AU \citep{2001AJ....121.2804B}. 
Subsequent observations revealed that these objects
shared a common red coloring \citep{2002ApJ...566L.125T}. 
In the H/WTSOSS survey these objects do not fall along mixing
lines, as the neutral and red clumps of centaurs do, but instead appear
nearly uniform in color space \citep{hwtsoss}.
These cold classical KBOs are
also know to be unique in their lack of large bodies 
\citep{2001AJ....121.1730L}, their
higher abundance of satellites \citep{2008Icar..194..758N}, 
and their different size distribution \citep{2010Icar..210..944F}.
And though our understanding of albedos in the Kuiper belt is
still poor, preliminary results suggest that the cold classical KBOs also appear to have higher albedos than
those of the remaining population
\citep{2009Icar..201..284B}. 
All of these properties appear to signify a population with a 
different -- and perhaps unique -- formation location or history. 
Understanding the surface compositions of these objects will be 
a challenge given their distance and small sizes. In addition, these
objects are dynamically stable, so we likely never get samples
of this population as centaurs or comets.

\subsection{The diverse colors of the remainder of the Kuiper belt}
Other than the unique color properties of the centaurs and cold classical
KBOs, the bulk of KBOs have no discernible pattern to their colors.
This finding in itself is significant for understanding the causes
of the colors of KBOs.
This lack of
a connection between color and any dynamical property, 
particularly with semi-major axis or perihelion/aphelion 
distance argues strongly that local heating, UV irradiation, 
and solar wind and cosmic ray
bombardment \citep{2003EM&P...92..261C} cannot be responsible for
the varying colors of the Kuiper belt. Local conditions appear to
have no primary influence on the colors of KBOs. 
Furthermore, careful measurement of the colors of the separate 
components of binary KBOs has shown a tight correlation over the full
range of Kuiper belt colors \citep{2009Icar..200..292B}. 
The colors of two KBOs in orbit around
each other are almost always nearly identical.
This fact immediately rules out any of the stochastic process such as
collisions for the causes of these Kuiper belt colors. Indeed, 
given the lack of correlation of color with local conditions, the
nearly identical colors of binary KBOs argues that colors are simply primordial.
If binary KBOs were formed by early mutual capture in a 
quiescent disk \citep{2002Natur.420..643G}, 
the two component would likely have formed in very similar locations. If,
alternatively, binary KBOs were formed in an initial gravitational
collapse \citep{2010AJ....140..785N}, the objects would of necessity have formed at the 
same location and of the same materials. 

\subsection{The transition from KBOs to centaurs}
The manner in which primordial KBO surfaces evolve to become
the color-bifurcated centaur population could provide important 
clues to the compositions of both surfaces. While it appears
that a transition from a unimodal to bimodal color distribution
must occur as objects move to lower perihelia, the actual
evidence for change is, in fact, weak.
The KBOs and centaurs with measured colors
have very different ranges of sizes, 
and, as demonstrated above, large KBOs have their
surfaces modified by the presence of volatiles and water ice. We
thus must only compare like-sized objects. In addition, the cold
classical KBOs, with their unique surfaces, likely never enter the
centaur population, so these should be excluded from the 
comparison. Finally,
the KBOs, being more distant, are likely to have higher uncertainty in
their color measurements. 

\begin{figure}
\plotone{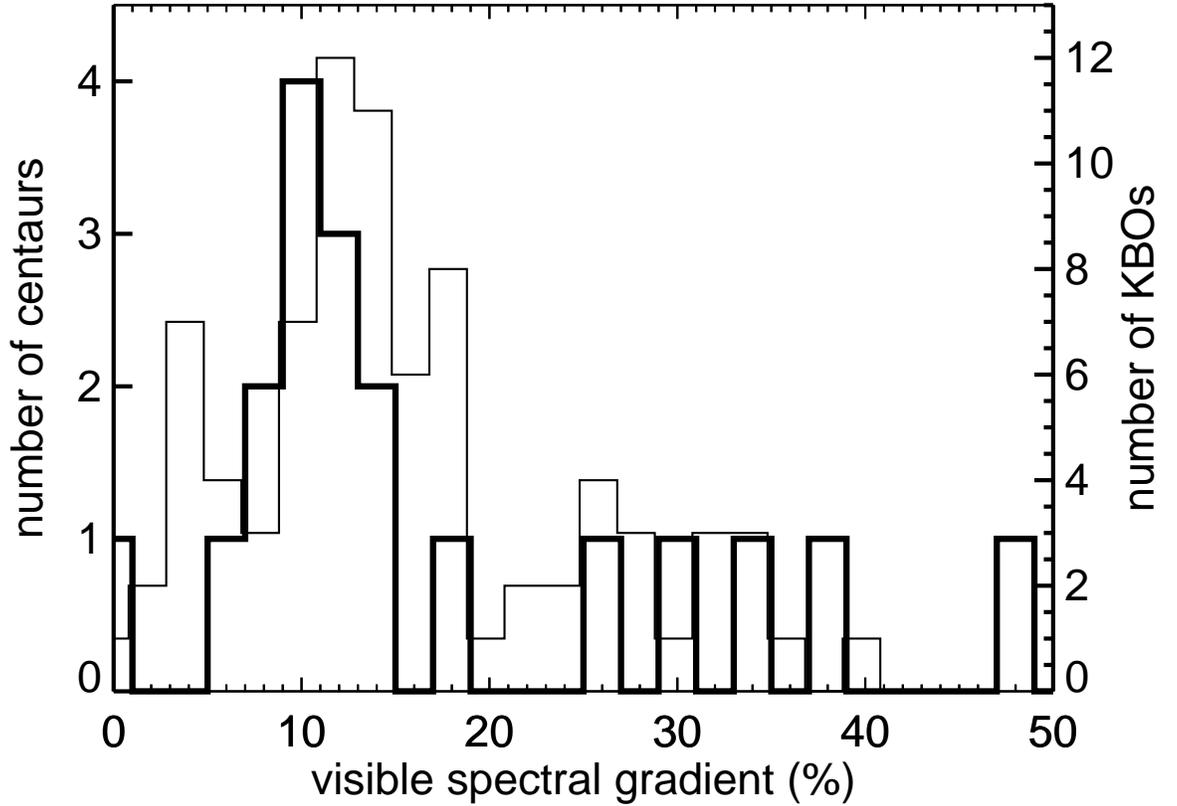}
\caption{
A comparison of the colors of (non-cold classical)
KBOs (thin line) and centaurs (thick line) in the $6<H<9$ range.
A K-S test cannot distinguish any significant difference between
the distributions. While it is commonly assumed that the surfaces
of KBOs evolve to have the color distribution of the centaur
population, we find no statistical evidence that such evidence
occurs. A similar conclusion can be drawn from the spectra
of Figure 5 and from the three-color data of the H/WTSOSS survey.}
\end{figure}

When the centaurs are compared to the appropriate KBOs, little evidence
of surface evolution can be found. 
In Figure 8 we compare well-measured optical colors of centaurs with absolute
magnitudes between 6 and 9
 to those of non-cold classical KBOs with the same range
of absolute magnitudes. While colors of the Kuiper belt as a whole 
appear significantly different from those of the centaurs, the 
difference is much less apparent when we compare the correct samples.
In fact,
a Kolmogorov-Smirnov test of the two distributions cannot distinguish the two at
greater than a 48\% confidence limit. 
Using all of the available color data, we find
no statistical evidence that any color evolution has occurred
as objects move inward from the Kuiper belt to become centaurs.
\citet{hwtsoss} come to the same conclusion from the optical and near-infrared
photometry of the H/WTSOSS survey. In addition, the albedos of KBOs measured
with Spitzer appear to be similarly distributed to those of the
smaller KBOs \citep{2008ssbn.book..161S}.

While there is no evidence in the spectroscopic or photometric data
for surface evolution in centaurs, in at least one way centaurs and
KBOs are clearly different. A small number of centaurs is known to
show cometary activity, at least sporadically \citep{2009AJ....137.4296J}. 
To date,
such activity has appeared confined to objects in the neutral 
clump, though the numbers remain small and the color measurements themselves
could be compromised by the presence of a coma \citep{2009AJ....137.4296J}. 
In addition, all but 2 of the 13 of the known active centaurs have perihelia
inside of 10 AU.
It has been speculated that outgassing could lead to surface 
modification and even to the bifurcated optical colors of the centaurs
\citep{2008ssbn.book..105T}.
Interestingly, an examination of the 16 centaurs observed in the
H/WTSOSS survey
shows no evidence for activity in any of them. Only
4 of the 16 have perihelia inside of 10 AU, though. Nonetheless,
the colors bifurcation seen in this sample of currently inactive
centaurs suggests that color bifurcation is not directly
caused by centaur activity.

Robustly determining whether or not centaur surfaces are evolved
is difficult. The numbers of centaurs available is small, the
comparison KBOs are faint, and the expected distributions are 
unknown. It is possible that a complete understanding of this
question will require large scale surveys like that of 
the currently-proposed
Large Scale Synoptic Telescope (LSST)
to find and characterize
many more objects and acquire sufficient statistics to understand 
these populations.

\subsection{The causes of colors}
To date, only three hypotheses have been advanced to explain the colors
of KBOs. 
The earliest hypotheses suggested
randomized collisional excavation
 \citep{1996AJ....112.2310L} or velocity dependent impact resurfacing 
\citep{2002AJ....124.2297S}, but the correlated colors
of KBO binaries \citep{2009Icar..200..292B} effectively rules out these or any other stochastic
processes.  \citet{2003EM&P...92..261C} have suggested that the uniform red
colors of the cold classical KBOs are caused by their existence
in an irradiation minimum environment, with irradiation increasing interior
due to increasing fluxes from solar energetic ions and increasing
exterior by increased flux of energetic ions from the termination shock,
but this hypothesis has no general explanation for the remaining colors,
nor would it lead to the sizable observed population
neutrally colored higher
inclination objects in the same environment. 

A new hypothesis has suggested that the colors of KBOs are set by
early evaporation and irradiation of volatiles followed by dynamical
mixing \citep{2011ApJ...739L..60B}.
In this hypothesis, objects are formed outside of Neptune
in the $\sim$15-30 AU region in the early solar system out
of a diverse and variable mix of materials. When the nebula disperses
and objects are first exposed to sunlight, their surfaces heat, and
volatiles are driven off. For widely variable starting compositions,
the surface compositions of objects at the same distance from the sun
will become nearly identical. Figure 9 shows a model from
\citet{2011ApJ...739L..60B} which shows the distance at which different volatiles would be
completely removed from the surface of an object as a function of size
of the object. Regardless of the diversity of initial compositions, 
the surfaces of the objects will quickly have strong gradients in 
volatile composition. These volatiles are then irradiated and 
begin to develop the colors expected from their particular mix of
remaining surface volatiles.
\begin{figure}
\plotone{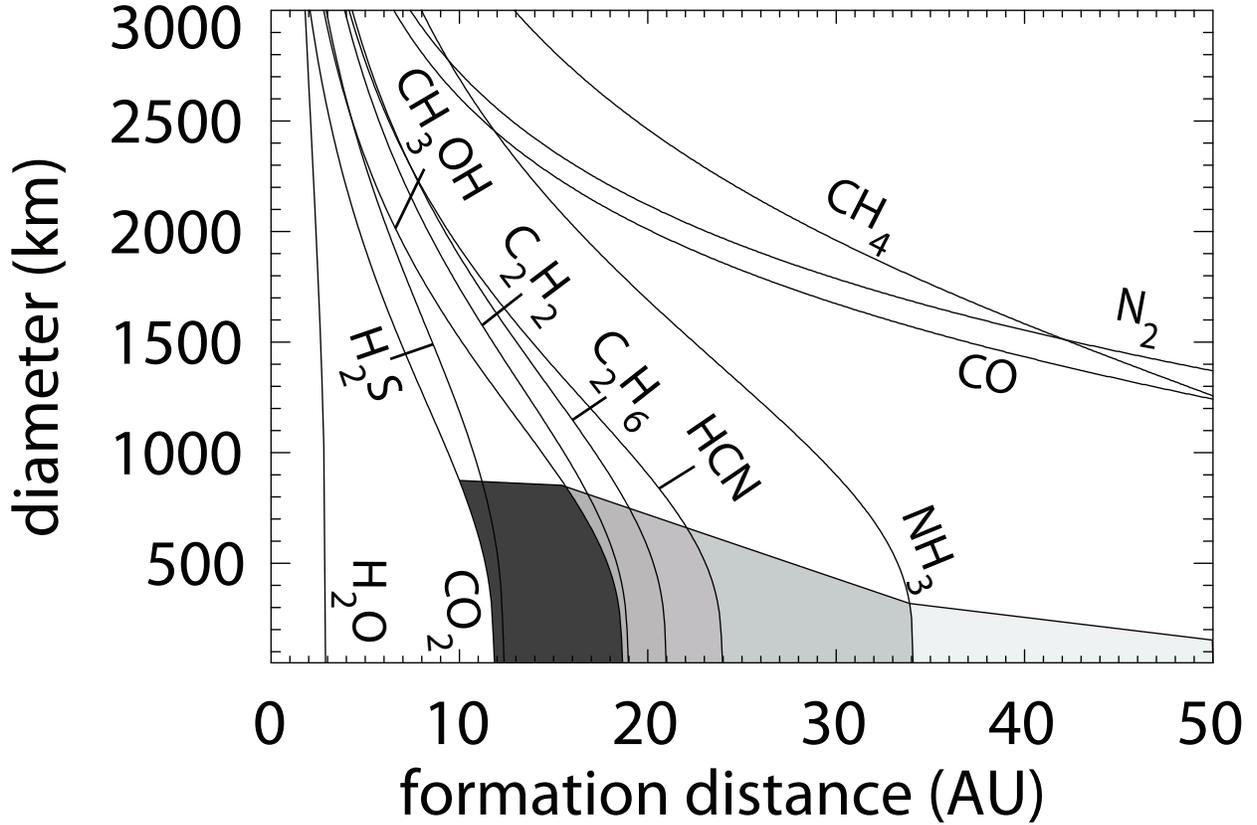}
\caption{
A hypothesis for the surface colors of small KBOs. In the early solar
system KBOs form with a variety of composition, but as the nebula
disappears and the sun begins to heat their surfaces, strong
gradients in their surface compositions form. Leftward of each
labeled line, each species would be fully depleted from the surface.
KBOs formed interior to about 20 AU would only have H$_2$O and CO$_2$
on their surfaces, 
while KBOs which formed outside of $\sim$20 AU
would retain methanol on their surfaces. Irradiation of the methanol-free
surfaces could lead to the dark neutral objects, while irradiation of
methanol-containing surfaces could lead to the brighter red objects.
Outside of $\sim$30 AU objects can also retain NH$_3$, which could
perhaps explain some of the unique surface characteristics of the
cold classical KBOs if they formed {\it in situ}.
}
\end{figure}

In this model, methanol is the most important 
coloring agent. Methanol is depleted on all surfaces inside of
about 20 AU, and present on all of those exterior.
While hydrocarbons are generally expected to carbonize and 
turn dark and neutral upon prolonged irradiation, 
laboratory experiments have suggested that irradiation of
methanol to dosages expected on KBOs over the age of
the solar system leads to bright red surfaces.
In this hypothesis, then,  the objects interior to the methanol line
become the neutral population, while those outside become the red
population. A solar system scale instability such as that envisioned by
the Nice model then scatters the objects onto the orbits where they
currently reside, mixing the neutral and red populations. The cold classical
KBOs form in place \citep{2012ApJ...744L...3B}, and their unique colors form by the
additional presence of ammonia on their surfaces. 
As KBOs scatter inward and become centaurs, their surfaces remain unchanged.

While the model of \citet{2011ApJ...739L..60B} is the only quantitative model to attempt to
explain all of the current observations of colors of KBOs, it is
clearly speculative. Much of the laboratory data required to trace
the specific irradiation chemistry has not been performed, and the question
of the evolution of the surfaces of the centaurs remains open. 
Extending this model to the Hubble Space Telescope survey of \citet{hwtsoss}, which suggests mixing
on the surfaces of KBOs, we would conclude that the irradiated volatiles
are the neutral and red materials that are mixed in with the hydrated
silicate-like material. No obvious explanation exists for the
different amounts of mixing between the irradiated volatiles
and the silicates.

\section{Bulk composition}
While most work on the composition of the Kuiper belt has focused on
the surface composition, all of the spectral and photometric results 
discussed probe an insignificantly small depth into the surface. To
understand the true composition of KBOs requires an understanding
of the bulk composition. Detailed measurement of the bulk
composition is of course impossible, but one important proxy --
the ice-to-rock fraction -- is available for some KBOs. Measurement
of the ice-to-rock ratio requires measurement of the density which, in
turn, requires measurement of the mass and radius of the objects. 
While measurement of the size of an object is possible through multiple 
means (to date far infrared radiometry from the Spitzer Space Telescope
has been the dominant method), a measurement of the mass is only possible
if the object has a satellite whose orbit is known.

KBOs
were expected to be a relatively homogeneous group. 
They all are thought to have
grown gradually through accretion, 
sampling similar regions of the solar nebula, so their compositions
should have been nearly identical. Indeed, when Pluto
was the only known large Kuiper belt object, its density of 
$\sim$2 g cm$^{-3}$ was taken to indicate a $\sim$30-70 ice-rock 
mix in the outer solar nebula as a whole \citep{1988Natur.335..240M}.

\begin{figure}
\plotone{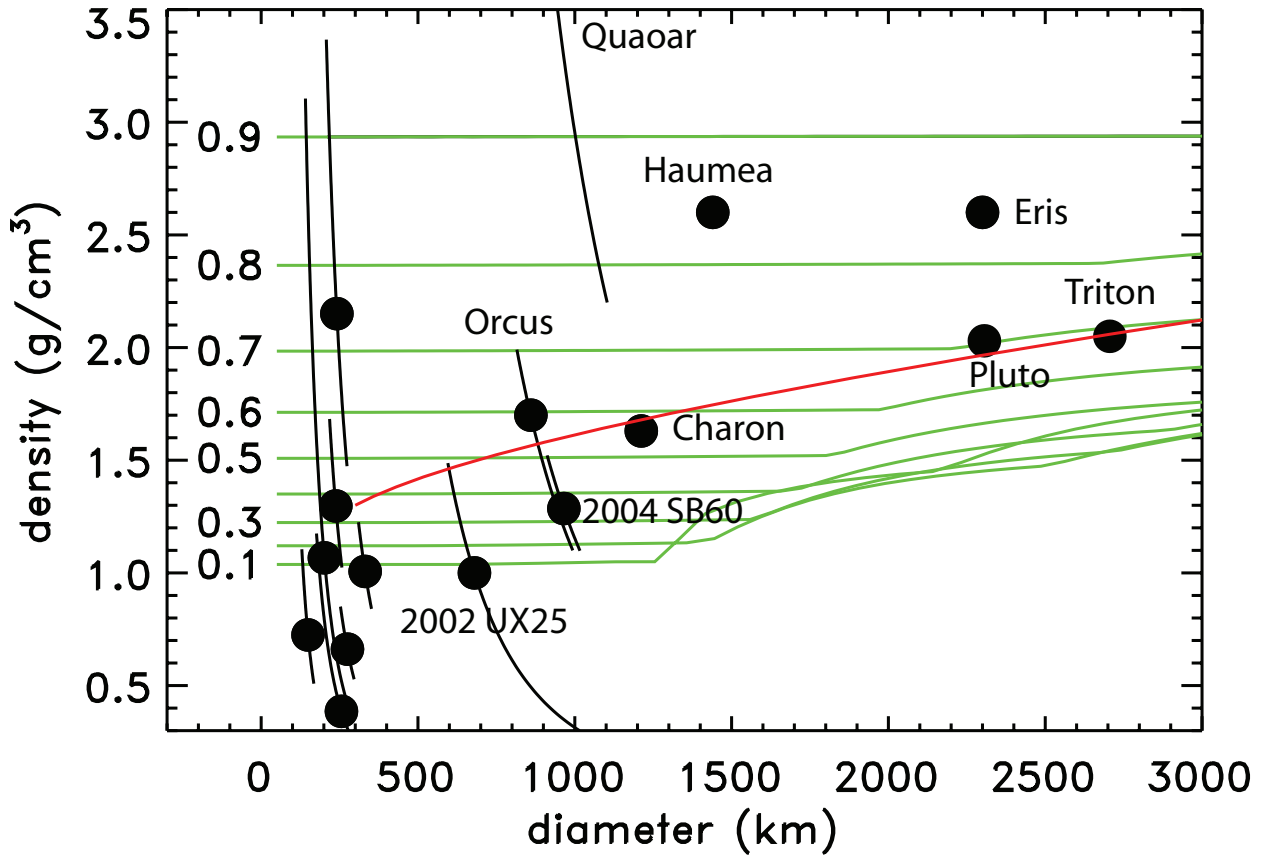}
\caption{Measured densities of KBOs. In all cases the uncertainty
in the size dominates the density measurement, so error bars follow
lines of constant mass with varying diameter. The green lines show
the density as a function of size for different rock fractions for
fully differentiated body; the ice phase change at higher
pressures causes ice to get denser. The red line shows the maximum
density increase that could be expected from a simple model where
small amounts of ice are preferentially removed in the collisions
required to build larger bodies from smaller ones. 
Data for sizes and densities come from \citet{2006ApJ...639.1238R,
2007Sci...316.1585B, 
2009DPS....41.6204M,
2010AJ....139.2700B, 
2010ApJ...714.1547F,2010Icar..207..978B}.
}
\end{figure}
One of the biggest outer solar system surprises of the past few years,
therefore, has been the discovery that the ice fraction
measured in KBOs varies from essentially 0 to 1 (Figure 10). Objects
have been found with densities significantly less than 1 g cm$^{-3}$
\citep{2009DPS....41.6204M,2010Icar..207..978B}
indicating both a near-unity ice fraction and significant porosity,
while other objects have been found with densities of nearly
pure rock \citep{2010ApJ...714.1547F}. 

Examination of the measured densities 
diameter reveals 
that the density measurements have extremely large
uncertainties. The large error bars are a function of the 
uncertainty in the measured sizes and the 3-times-higher uncertainties
in the associated volumes. Detailed understanding of the trends and
causes of ice-rock fractions in the Kuiper belt clearly requires 
significantly higher quality size measurements.

Even with these large uncertainties, however, two trends are apparent.
First, there is a general trend for an increase in density as a function 
of size. While an object with a fixed ice fraction will 
undergo a density increase with increased size owing to
the  small density increase that occurs due to the change 
of ice to higher density phase as pressure is increased, 
the general trend of increased density with size seen in the Kuiper
belt is significantly larger than expected unless
the rock fraction itself is increasing in larger objects (Fig. 10). 

The second general trend is the difference in densities between objects
which exclusively have small satellites and those which have larger satellites.
The objects which exclusively have
small satellites -- Haumea, Quaoar, and Eris -- have been hypothesized
to have undergone giant impacts which led to these satellite \citep{2006ApJ...639L..43B}.
Interestingly, these objects have higher densities than every other
measured object.

Such a wide range of ice-rock ratios is astounding.
No dynamical evidence exists that the large KBOs came from dramatically
different regions of the solar nebula, and
no reasonable formation mechanism appears capable of delivering
such variability among large accreted objects. 
We suggest three scenarios for explaining the extreme variability of
the bulk composition of KBOs below, but we do not find any 
of the explanations completely satisfying.

\subsection{Single collision densification}
The extreme densities of all of the KBOs which exclusively have
small satellites immediately suggests a connection between the
satellite-forming giant impact and the high density. Indeed, in
the case of Haumea, the creation of an icy collisional family
dynamically surrounding the object demonstrates that these
impacts do indeed remove ice from a differentiated mantle. 
It appears reasonable to suggest that the three known objects
with the highest densities all achieved these densities 
through single catastrophic collisions which removed 
substantial amounts of overlying ice in a process reminiscent of
that envisioned for Mercury.

Unfortunately, the types of collisions that would be required
are infeasible in the solar system. Using scalings from numerical
simulations of \citet{2009ApJ...691L.133S}, we find that to 
remove enough ice mass to create 
an object with the $\sim$2.6 g cm$^{-3}$ of Eris from an object that initially had
the $\sim$2.0 g cm$^{-3}$ of Pluto requires hitting Triton with
a $\sim$ 500 km KBO traveling 40 km s$^{-1}$! 

Though the connection between high densities and small satellites
is striking, we can envision no physical mechanism by which the
two can be related.

\subsection{Water ice loss during accretion}
While single impacts cannot lead to significant ice loss, 
multiple impacts during accretion can.
The red line in Figure 10 shows, for example,
a scenario where large bodies are preferentially grown by
accreting other like-sized bodies. For two bodies which collide
at their mutual escape velocities, as such bodies would, 
the \citet{2009ApJ...691L.133S} scalings suggest that $\sim$10\%
of the total mass will be removed from the system. If for 
differentiated objects the ice is exclusively removed, these objects
will grow in density as they grow in size. 

The model shown here
then represent the maximum density 
growth that could happen from an assumed starting
size and starting density (300 km, 1.3 g cm$^{-3}$)
for differentiated objects which grow to 
become the larger dwarf planets. While a general increase such as that
seen through Charon, Pluto, and Triton can perhaps be explained,
the three very high density objects are well away from this line.
To create Quaoar, for example, would require starting with 300 km 
differentiated objects
with initial densities of 1.9 g cm$^{-3}$, which would then be
inconsistent with the other observed densities.

While accretional loss of water ice may play some role in the 
general trend of increasing densities with size, that process
clearly cannot explain all of the observations.

\subsection{Extreme inhomogeneity in the disk}
One of the only alternatives to invoking collisions as a means to 
modify densities is to suggest that the densities are primordial.
The general trend of densities increasing with size could be caused
by an increase in the ice-to-rock ratio as a function of heliocentric
distance, combined by a tendency of the largest objects to form
closer to the sun where formation timescales are shorter and
larger objects can grow. However, this scenario would also 
suggest the existence of a population of smaller objects
which also formed at these closer distances and should therefore
also have the high densities of the largest objects. No such
population appears. 

The alternative is that the ice-to-rock ratio did not vary
smoothly with distance, but rather the disk was chemically 
inhomogeneous. It is difficult to reconcile this possibility
with the assumption that the large objects were accreted out
of large numbers of small ones sampling multiple portions of the disk.

\bigskip
None of the three proposed scenarios provides a satisfactory 
explanation for the trends and variability seen in densities.
The study of the bulk composition of KBOs is, however,
 in its infancy. Radii of KBOs are still poorly determined, and
the number of objects with measured satellite orbits remains small.
We anticipate that continued work in this areas will eventually 
yield the insights that will allow the bulk compositions
of these objects to be understood.

\section{Implications, speculations, and open questions}
We have attempted to discern general trends and formulate
overlying principles for understanding the compositions of the
objects in the 
Kuiper belt. We are at a stage in our understanding of the 
Kuiper belt that, for the first time, we have at least a first-order
understanding of the compositions of the surfaces
of KBOs and valid working hypotheses for understanding these compositions.
A comparable understanding of the bulk composition is -- however --
far away.

Using all of the spectroscopic and photometric data as well as 
the current best understanding of the physics and chemistry of these
bodies, 
we conceptually divide Kuiper belt surfaces into major groups, 
with differences being caused by size, formation location, and history.
The six broad types of KBO surfaces are as follow:
\begin{itemize}
\item{Volatile rich (Triton, Eris, Pluto): objects large enough to 
retain significant reservoirs of N$_2$, CH$_4$, and CO. Their spectra
are dominated by CH$_4$ absorption, but their surfaces 
are likely dominated by N$_2$ with CH$_4$ a moderate contributor.}
\item{Volatile transition (Makemake, Quaoar, 2007 OR10): objects which 
are on the verge of losing their volatiles but still contain (at least)
CH$_4$. With CH$_4$ the dominant molecule on the surface, radiation
processing occurs much more quickly, so ethane and other methane irradiation
products are present on the surface.}
\item{Water ice plus ammonia rich surfaces (Charon, Orcus, Quaoar, 
2007 OR10, AZ84, other): Slightly smaller ($\sim 500$ km $< D < \sim$1200 km)
objects on which volatiles have been mostly
or fully depleted and the spectra are dominated by significant crystalline
water ice absorption. Where high S/N is available and methane doesn't
hide the spectral signature, we expect that all of these will show
the presence of ammonia. We suggest that these are due to water
ice flowing on the surface after volatile irradiation has otherwise
set the surface color and composition.}
\item{Neutral surfaces of small objects:} Objects which are too small to
retain volatiles or to have had water flows and formed inside $\sim$20 AU
will have their surfaces depleted in all major ices except H$_2$O and CO$_2$.
Irradiation will then cause dark neutrally colored surfaces to develop.
These objects show a range of optical-near IR colors based on the
amount of silicate mixed with the irradiated ices. No hypothesis has
been suggested for why objects differ in their surface rock-ice ratio.
\item{Red surfaces of small objects:} Small objects which formed outside
of $\sim$20 AU would have been able to retain CH$_3$OH on their surface. Upon
irradiation these surfaces would turn red and retain moderate albedos.
These objects show a range of optical and optical-near IR colors based
on the amount of silicate mixed with the irradiated ices. 
\item{Pure water ice (Haumea and its family and satellites): 
pure water ice surfaces
form when the nearly-pristine water ice mantle of a differentiated KBO
is exposed in an impact.}
\end{itemize}

While these six classes of Kuiper belt surfaces provide an overall framework
for understanding the the composition of KBO surfaces, there are 
open questions about multiple aspects of this framework that must
be answered before we can be certain that our first-order understanding
of these surfaces is correct. In addition, some overall questions still
remain. We summarize some of the most important questions to be
answered by future research:

\begin{itemize}
\item{Is water ice on medium-sized KBOs a sign of previous water flow on
the surface? Evidence that this hypothesis is true would include
the presence of ammonia on every medium-sized objects on which substantial
water ice absorption is present. Images of Charon from the New Horizon
spacecraft may also shed light on this question.}
\item{Is the sporadic methanol seen on small objects a transient sign
of a recent impact? Direct evidence for this hypothesis is difficult
without resolved imaging, for which there are no future plans. However,
rotationally revolved spectroscopy could help begin
to answer this question. Such spectroscopy will require the next
generation of large telescopes, but rotationally resolved color measurement
may provide a currently feasible initial step.}
\item{Do KBO surfaces become modified as they enter the centaur region? 
Extremely high precision color measurements of large numbers of 
higher perihelion non-cold classical KBOs
could answer this question. Such precise measurements are possible
with current telescopes. Understanding the conditions that lead to
centaur activity and that effect of activity on the surface -- if any --
will also be important.}
\item{Can the volatile loss hypothesis for the colors of KBOs be
supported by new laboratory data? The hypothesis suggests specific 
chemical combinations that give the colors seen in the Kuiper belt;
irradiation of these combinations would show if the predictions
are valid.}
\item{Are silicates visible at the surface? Reliably reproducible 
results on the presence of aqueously altered silicates would
be strong evidence. Midinfrared spectroscopy is difficult but
should eventually prove feasible. The silicate-colored
material identified in the mixtures from the purely
photometric H/WTSOSS surface could perhaps be isolated 
spectroscopically with carefully targeted measurements.}

\item{What causes the extreme bulk
compositional variability of the Kuiper belt? Three hypotheses have
been suggested, but none of them is satisfactory. Significant more
work is needed in this area.}
\end{itemize}

The last decade has seen studies of the composition
of the Kuiper belt blossom from simple cataloging to 
the development of actual conceptual frameworks which
can explain many of the observed properties. Much work remains, however,
to both verify and extend this framework, and to use these
distant bodies to continue to provide new insights into the formation
and evolution of the solar system.

The preparation of this review has been supported by grant
NNX09AB49G from the NASA Planetary Astronomy program.

\end{document}